\documentclass[a4paper,11pt]{article}
\usepackage{pos}
\usepackage{amsmath}
\usepackage{dsfont}
\usepackage{bbm}
\usepackage{cleveref}
\usepackage{subcaption}
\usepackage[utf8]{inputenc}

\let\OLDthebibliography\thebibliography
\renewcommand\thebibliography[1]{
  \OLDthebibliography{#1}
  \setlength{\parskip}{0pt}
  \setlength{\itemsep}{0pt plus 0.3ex}
}

\DeclareMathOperator{\str}{str}

\newcommand{\I}{\mathrm{i}}

\newcommand{\ren}{\mathrm{R}}

\newcommand{\upq}{\mathrm{u}}
\newcommand{\downq}{\mathrm{d}}
\newcommand{\strangeq}{\mathrm{s}}

\newcommand{\dmu}{\Delta m_{\upq}}
\newcommand{\dmd}{\Delta m_{\downq}}
\newcommand{\dms}{\Delta m_{\strangeq}}

\newcommand{\ee}{e^{2}}
\newcommand{\MeV}{\text{MeV}}

\newcommand{\fm}{\text{fm}}
\newcommand{\dif}{\mathrm{d}}
\newcommand{\phys}{\mathrm{phys}}

\newcommand{\qcdqed}{\mathrm{QCD+QED}}
\newcommand{\qcdiso}{\mathrm{QCD}_{\mathrm{iso}}}

\title{Leading isospin breaking effects in the HVP contribution to $a_{\mu}$ and to the running of $\alpha$}
\ShortTitle{Leading isospin breaking effects in the HVP contribution to $a_{\mu}$ and to the running of $\alpha$}

\author*[a]{Andreas Risch}
\author[b,c,d,e]{Hartmut Wittig}

\affiliation[a]{John von Neumann-Institut f{\"u}r Computing NIC, Deutsches Elektronen-Synchrotron DESY,\\
Platanenallee 6, 15738 Zeuthen, Germany}
\affiliation[b]{PRISMA$^{+}$ Cluster of Excellence, Johannes-Gutenberg-Universit{\"a}t Mainz,\\
Staudingerweg 9, 55128 Mainz, Germany}
\affiliation[c]{Institut f{\"u}r Kernphysik, Johannes-Gutenberg-Universit{\"a}t Mainz,\\
Johann-Joachim-Becher-Weg 45, 55128 Mainz, Germany}
\affiliation[d]{Helmholtz-Institut Mainz, Johannes-Gutenberg Universit{\"a}t Mainz,\\
Staudingerweg 18, 55128 Mainz, Germany}
\affiliation[e]{GSI Helmholtzzentrum f{\"u}r Schwerionenforschung,\\
Planckstra{\ss}e 1, 64291 Darmstadt, Germany}

\emailAdd{andreas.risch@desy.de}
\emailAdd{hartmut.wittig@uni-mainz.de}

\abstract{The anomalous magnetic moment of the muon $a_{\mu}$ and the running of the electromagnetic coupling $\alpha$ play a fundamental role in beyond Standard Model (SM) physics searches. Non-perturbative hadronic contributions to both quantities, which are related to the hadronic vacuum polarization (HVP) function consisting of two electromagnetic currents, are a main source of uncertainty in the SM prediction. We compute the HVP function in lattice QCD+QED applying the time-momentum representation method. We expand the relevant correlation functions around the isosymmetric limit. In particular, we focus on leading isospin breaking effects taking quark-connected contributions into account, which we evaluate on isosymmetric $N_{\mathrm{f}}=2+1$ QCD gauge ensembles generated by the CLS initiative with non-perturbatively $O(a)$-improved Wilson fermions.}

\FullConference{%
 The 38th International Symposium on Lattice Field Theory, LATTICE2021
  26th-30th July, 2021
  Zoom/Gather@Massachusetts Institute of Technology
}

\begin{document}
\maketitle

\section{Introduction}

The anomalous magnetic moment of the muon $a_{\mu}$ and the running of the electromagnetic coupling $\alpha$ play a fundamental role in searches for physics beyond the Standard Model (SM). For both quantities the hadronic vacuum polarisation (HVP) contribution is a main source of uncertainty in the SM prediction. In particular, at the desired level of accuracy isospin breaking effects in the HVP contribution have to be taken into account~\cite{Gerardin:2020gpp,Aoyama:2020ynm}. In this work, we continue the investigation of isospin breaking effects~\cite{Risch:2017xxe,Risch:2018ozp,Risch:2019xio} making use of Coordinated Lattice Simulations (CLS) $N_{\mathrm{f}}=2+1$ QCD ensembles~\cite{Bruno:2014jqa,Bruno:2016plf,Mohler:2017wnb,Mohler:2020txx} with open and (anti-)periodic temporal boundary conditions~\cite{Luscher:2011kk}.

We organise this work as follows: We briefly summarise the setup used for the perturbative treatment of isospin breaking effects~\cite{deDivitiis:2011eh, deDivitiis:2013xla} based on reweighting QCD$_{\text{iso}}$ gauge ensembles and discuss suitable hadronic renormalisation schemes for QCD+QED and QCD$_{\text{iso}}$ inspired by chiral perturbation theory. We recap the formalism for the computation of mesonic two-point functions and for the renormalisation of the local vector current in this framework using QED$_{\mathrm{L}}$~\cite{Hayakawa:2008an} as a finite-volume prescription of QED. We finally discuss isospin breaking effects in the LO-HVP contribution to the muon anomalous magnetic moment as well as in the closely related LO hadronic contributions to the running of the electromagnetic coupling.

\section{Inclusion of perturbative isospin breaking effects by reweighting}

We briefly summarise our setup for the perturbative treatment of isospin breaking effects. For a detailed description we refer to~\cite{Risch:2018ozp}. We consider the space of QCD+QED-like theories parameterised by $\varepsilon = (am_{\mathrm{u}}, am_{\mathrm{d}}, am_{\mathrm{s}},\beta, e^{2})$. For the choice $\varepsilon^{(0)} = (am_{\mathrm{u}}^{(0)}, am_{\mathrm{d}}^{(0)}, am_{\mathrm{s}}^{(0)}, \beta^{(0)}, 0)$ with $am_{\mathrm{u}}^{(0)}= am_{\mathrm{d}}^{(0)}$ we obtain QCD$_{\mathrm{iso}}$ together with a free photon field. In~\cite{Risch:2018ozp} we have shown that QCD+QED can be related to QCD$_{\mathrm{iso}}$ by reweighting via the identity
\begin{align}
\langle O[U,A,\Psi,\overline{\Psi}] \rangle &= \frac{\langle R[U] \,\langle O[U,A,\Psi,\overline{\Psi}] \rangle_{\mathrm{q}\gamma} \rangle_{\mathrm{eff}}^{(0)}}{\langle R[U] \rangle_{\mathrm{eff}}^{(0)}} & R[U] &= \frac{\exp(-S_{\mathrm{g}}[U])\,Z_{\mathrm{q}\gamma}[U]}{\exp(-S_{\mathrm{g}}^{(0)}[U])\,Z^{(0)}_{\mathrm{q}}[U]},
\label{eq_expectation_value_by_reweighting}
\end{align}
where $\langle \ldots \rangle_{\mathrm{eff}}^{(0)}$ is evaluated by making use of existing QCD$_{\mathrm{iso}}$ gauge configurations. $\left\langle \ldots \right\rangle_{\mathrm{q}\gamma}$ denotes the QED expectation value on a QCD background gauge field and $Z_{\mathrm{q}\gamma}[U]$ is the corresponding partition function, whereas $Z_{\mathrm{q}}^{(0)}[U]$ denotes the partition function of isosymmetric quarks on a QCD background gauge field, i.e. the quark determinant of QCD$_{\mathrm{iso}}$. We evaluate $R[U]$ by means of perturbation theory in $\Delta\varepsilon=\varepsilon-\varepsilon^{(0)}$ around $\varepsilon^{(0)}$. The required Feynman rules are discussed in~\cite{Risch:2018ozp}. In order to fix the expansion coefficients $\Delta\varepsilon$ we make use of a suitable hadronic renormalisation scheme discussed in the next section.

\section{Hadronic renormalisation scheme for QCD+QED and QCD$_{\text{iso}}$}

Masses of pseudo-scalar mesons can be computed in chiral perturbation theory including the electromagnetic interaction. Defining the average light quark mass $\hat{m} = \frac{1}{2}(m_{\upq}+m_{\downq})$ and the $\pi^{0}$-$\eta$ mixing angle $\varepsilon = \frac{\sqrt{3}}{4}\frac{m_{\downq}-m_{\upq}}{m_{\strangeq}-\hat{m}}$ the lowest-order contribution to pseudo-scalar meson masses at $O(e^{2}p^{0})$ and $O(\varepsilon)$ are given by~\cite{Neufeld:1995mu}
\begin{align}
m_{\pi^{+}}^{2} &= 2B\hat{m}+2e^{2}ZF^{2}, & m_{K^{+}}^{2} &= B\Big((m_{\strangeq}+\hat{m})-\frac{2\varepsilon}{\sqrt{3}}(m_{\strangeq}-\hat{m})\Big)+2e^{2}ZF^{2}, \nonumber \\
m_{\pi^{0}}^{2} &= 2B\hat{m}, & m_{K^{0}}^{2} &= B\Big((m_{\strangeq}+\hat{m})+\frac{2\varepsilon}{\sqrt{3}}(m_{\strangeq}-\hat{m})\Big),
\end{align}
where $F$ is the pion decay constant in the chiral limit, $B$ the vacuum condensate parameter and $Z$ a dimensionless coupling constant. The linear combinations $m_{\pi^{0}}^{2} = B(m_{\upq}+m_{\downq})$, $m_{K^{+}}^{2}+m_{K^{0}}^{2}-m_{\pi^{+}}^{2} = 2Bm_{\strangeq}$ and $m_{K^{+}}^{2}-m_{K^{0}}^{2}-m_{\pi^{+}}^{2}+m_{\pi^{0}}^{2} = B(m_{\upq}-m_{\downq})$ serve as proxies for the average light quark mass, the strange quark mass and the light quark mass splitting. Making use of the fact that at leading order $\alpha_{\mathrm{em}}$ does not renormalise, i.e. $\alpha_{\mathrm{em}}=\frac{e^{2}}{4\pi}$, we use the above expressions to define a hadronic renormalisation scheme for QCD+QED:
\begin{align}
&(m_{\pi^{0}}^{2})^{\qcdqed} \!=\! (m_{\pi^{0}}^{2})^{\phys}, \quad\quad (m_{K^{+}}^{2} + m_{K^{0}}^{2} - m_{\pi^{+}}^{2})^{\qcdqed} \!=\! (m_{K^{+}}^{2} + m_{K^{0}}^{2} - m_{\pi^{+}}^{2})^{\phys}\label{eq:QCDQEDscheme}, \\
&(m_{K^{+}}^{2}-m_{K^{0}}^{2}-m_{\pi^{+}}^{2}+m_{\pi^{0}}^{2})^{\qcdqed} \!=\! (m_{K^{+}}^{2}-m_{K^{0}}^{2}-m_{\pi^{+}}^{2}+m_{\pi^{0}}^{2})^{\phys}, \, (\alpha_{\mathrm{em}})^{\qcdqed} \!=\! (\alpha_{\mathrm{em}})^{\phys}.\nonumber
\end{align}
The superscript "phys" indicates the experimentally measured value, whereas "QCD+QED" refers to the theoretical prediction. Forming appropriate linear combinations the above scheme is equivalent to matching the quantities $m_{\pi^{0}}^{2}$, $m_{K^{0}}^{2}$ and $m_{K^{+}}^{2} - m_{\pi^{+}}^{2}$ instead. Optionally, $m_{\pi^{+}}^{2}-m_{\pi^{0}}^{2} = 2e^{2}ZF^{2}$ can be used as a proxy for $\alpha_{\mathrm{em}}=\frac{e^{2}}{4\pi}$, such that one obtains a scheme based on $m_{\pi^{0}}^{2}$, $m_{\pi^{+}}^{2}$, $m_{K^{0}}^{2}$ and $m_{K^{+}}^{2}$. In addition, it is possible to introduce a scheme for QCD$_{\text{iso}}$, which is characterised by a vanishing electromagnetic coupling and identical up- and down-quark masses. In this case, in \cref{eq:QCDQEDscheme} only two proxies for the quark masses remain:
\begin{align}
\big(m_{\pi^{0}}^{2}\big)^{\qcdiso} &= \big(m_{\pi^{0}}^{2})^{\phys}, & \big(m_{K^{+}}^{2} + m_{K^{0}}^{2} - m_{\pi^{+}}^{2}\big)^{\qcdiso} &= (m_{K^{+}}^{2} + m_{K^{0}}^{2} - m_{\pi^{+}}^{2}\big)^{\phys}.
\end{align}
Combining the latter equations and making use of the fact that the pions and kaons become mass degenerate, respectively, one finds for the squared isosymmetric pion and kaon masses~\cite{Blum:discussion2021workshop}:
\begin{align}
\big(m_{\pi}^{2}\big)^{\qcdiso} &= \big(m_{\pi^{0}}^{2}\big)^{\phys}, & \big(m_{K}^{2}\big)^{\qcdiso} &= \frac{1}{2}\big(m_{K^{+}}^{2} + m_{K^{0}}^{2} - m_{\pi^{+}}^{2} + m_{\pi^{0}}^{2}\big)^{\phys}.
\end{align}
Isospin breaking effects of an observable $O$ can now be quantified by comparing the predictions $(O)^{\qcdqed}$ and $(O)^{\qcdiso}$. Similarly, a scheme for QCD is obtained when demanding a vanishing electromagnetic coupling in \cref{eq:QCDQEDscheme}. These chiral perturbation theory inspired schemes have the advantage to be purely based on pseudo-scalar meson masses and are therefore, in contrast to schemes based on renormalised quark masses~\cite{deDivitiis:2013xla}, easy to handle.

Since the limited number of gauge ensembles considered so far does not yet allow for an extrapolation to the physical point, we match QCD+QED and QCD$_{\mathrm{iso}}$ on each ensemble. Neglecting isospin breaking effects in the scale setting we match the proxies for the average light and strange quark masses in both theories and set the light quark mass difference and the electromagnetic coupling to their physical values. Applying the leading-order perturbative expansion $am_{H}=(am_{H})^{(0)}+\sum_{l}\Delta\varepsilon_{l}(am_{H})^{(1)}_{l}+O(\Delta\varepsilon^{2})$ for $H=\pi^{0},\pi^{+},K^{0},K^{+}$ this scheme translates into a system of linear equations that determines the expansion coefficients $\Delta\varepsilon=(a\Delta m_{\mathrm{u}}, a\Delta m_{\mathrm{d}}, a\Delta m_{\mathrm{s}},\Delta\beta, e^{2})$:
\begin{align}
&\sum_{l} \Delta\varepsilon_{l} \Big((am_{\pi^{0}})^{(0)} (am_{\pi^{0}})^{(1)}_{l}\Big) = 0, \quad\quad\quad\quad\quad\quad \Delta\varepsilon_{\Delta\beta} = 0, \quad\quad\quad\quad\quad\quad
\Delta\varepsilon_{e^{2}} = 4\pi\alpha_{\mathrm{em}}, \nonumber
\end{align}
\begin{align}
&\sum_{l} \Delta\varepsilon_{l} \Big((am_{K^{+}})^{(0)} (am_{K^{+}})^{(1)}_{l} + (am_{K^{0}})^{(0)} (am_{K^{0}})^{(1)}_{l} - (am_{\pi^{+}})^{(0)} (am_{\pi^{+}})^{(1)}_{l}\Big) = 0, \nonumber \\
&\sum_{l} \Delta\varepsilon_{l} \Big(
\begin{aligned}[t]
&(am_{K^{+}})^{(0)}(am_{K^{+}})^{(1)}_{l}-(am_{K^{0}})^{(0)} (am_{K^{0}})^{(1)}_{l}-(am_{\pi^{+}})^{(0)} (am_{\pi^{+}})^{(1)}_{l}\\
&+(am_{\pi^{0}})^{(0)}(am_{\pi^{0}})^{(1)}_{l}\Big) = \frac{1}{2} a^{(0)}\big(m_{K^{+}}^{2}-m_{K^{0}}^{2}-m_{\pi^{+}}^{2}+m_{\pi^{0}}^{2})^{\phys}.
\end{aligned}
\end{align}

\section{Mesonic two-point correlation function}

To determine pseudo-scalar meson masses required for the hadronic renormalisation scheme as well as to compute the HVP function we consider zero-momentum projected mesonic two-point functions for the operator combinations $(\mathcal{M}_{2},\mathcal{M}_{1})=(\mathcal{P},\mathcal{P}),(\mathcal{V}_{\mathrm{l}},\mathcal{V}_{\mathrm{l}}),(\mathcal{V}_{\mathrm{c}},\mathcal{V}_{\mathrm{l}})$:
\begin{align}
C(x_{2}^{0},x_{1}^{0}) &=\frac{a^{6}}{|\Lambda_{123}|}\sum_{\vec{x}_{2},\vec{x}_{1}}\langle \mathcal{M}_{2}^{x_{2}}\mathcal{M}_{1}^{x_{1}} \rangle = \langle \mathcal{M}_{2}^{x_{2}^{0}}\mathcal{M}_{1}^{x_{1}^{0}} \rangle &\text{with}\quad&\mathcal{M}^{x^{0}}_{i}=\frac{a^{3}}{\sqrt{|\Lambda_{123}|}}\sum_{\vec{x}}\mathcal{M}^{x}_{i}. \label{label1}
\end{align}
$|\Lambda_{123}|$ denotes the spatial volume of the lattice. The pseudo-scalar density operator is defined as $\mathcal{P}^{x i} = \overline{\Psi}{}^{x}\Lambda^{i}\gamma^{5}\Psi{}^{x}$, where $\Lambda^{i}$ determines the flavour content. We make use of two lattice discretisations of the vector current: the ultra-local discretisation $\mathcal{V}_{\mathrm{l}}^{x\mu i} = \overline{\Psi}{}^{x}\Lambda^{i}\gamma^{\mu}\Psi{}^{x}$ and the conserved discretisation $\mathcal{V}_{\mathrm{c}}^{x\mu i} = \frac{1}{2}\big(\overline{\Psi}{}^{x+a\hat{\mu}}(W^{x\mu})^{\dagger}\Lambda^{i}(\gamma^{\mu}+\mathds{1})\Psi{}^{x}+\overline{\Psi}{}^{x}\Lambda^{i}(\gamma^{\mu}-\mathds{1})W^{x\mu}\Psi{}^{x+a\hat{\mu}}\big)$~\cite{Risch:2019xio}, which fulfils the lattice vector Ward identity in QCD+QED for diagonal $\Lambda^{i}$ and which depends on the combined QCD+QED gauge links $W^{x\mu} = U^{x\mu} e^{\I a e Q A^{x\mu}}$. Treating isospin breaking perturbatively, correlation functions are expanded according to $C=C^{(0)}+\sum_{l}\Delta\varepsilon_{l}C^{(1)}_{l}+O(\Delta\varepsilon^{2})$. As a consequence, operators also have to be expanded in $e$, i.e. $\mathcal{O}=\mathcal{O}^{(0)}+e\mathcal{O}^{(\frac{1}{2})}+\frac{1}{2}e^{2}\mathcal{O}^{(1)}+O(e^{3})$. The expansion of $\mathcal{V}_{\mathrm{c}}$ in $e$ can be found in~\cite{Risch:2019xio}. Combining \cref{eq_expectation_value_by_reweighting} and \cref{label1}, the quark-connected 0th and 1st order contributions to the mesonic two-point functions read
\begin{align}
C{}^{(0)} &= \Big\langle
\begin{gathered}
\includegraphics[width=6.5em]{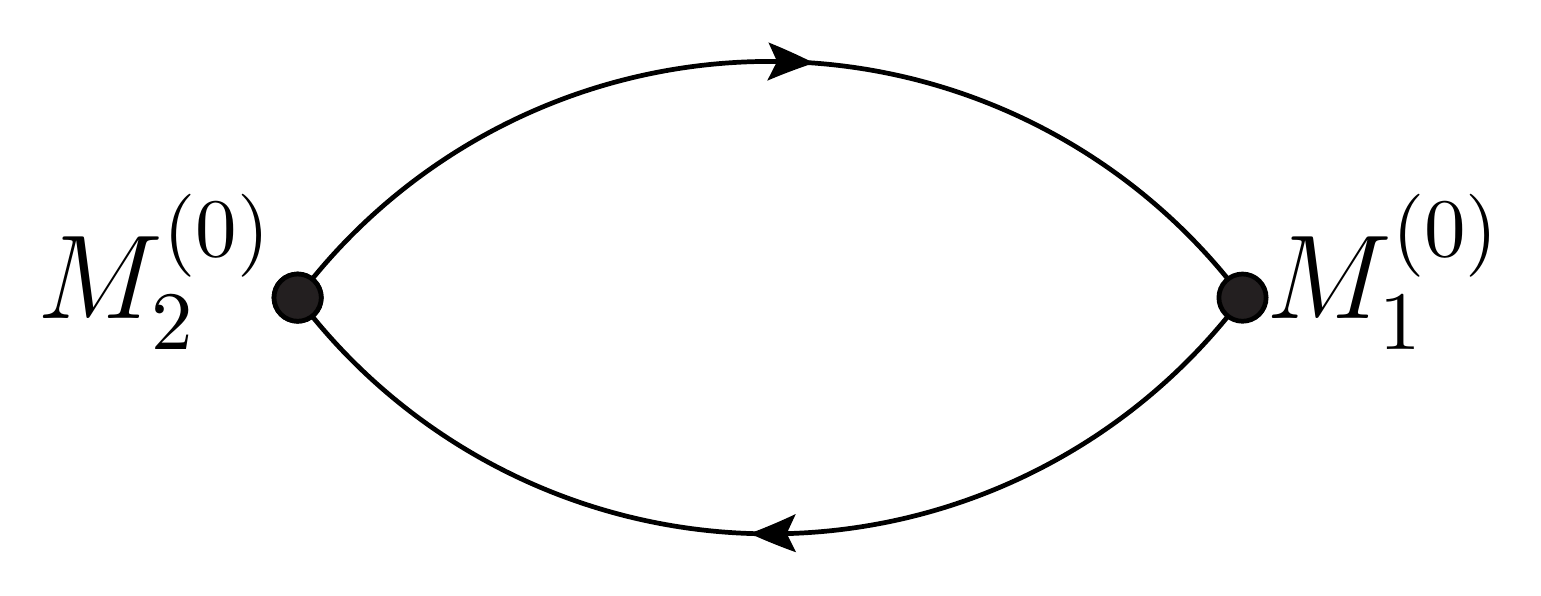}
\end{gathered}
\Big\rangle_{\mathrm{eff}}^{(0)},
\quad\quad\quad\quad C{}^{(1)}_{\Delta m_{f}} =
\Big\langle
\begin{gathered}
\includegraphics[width=6.5em]{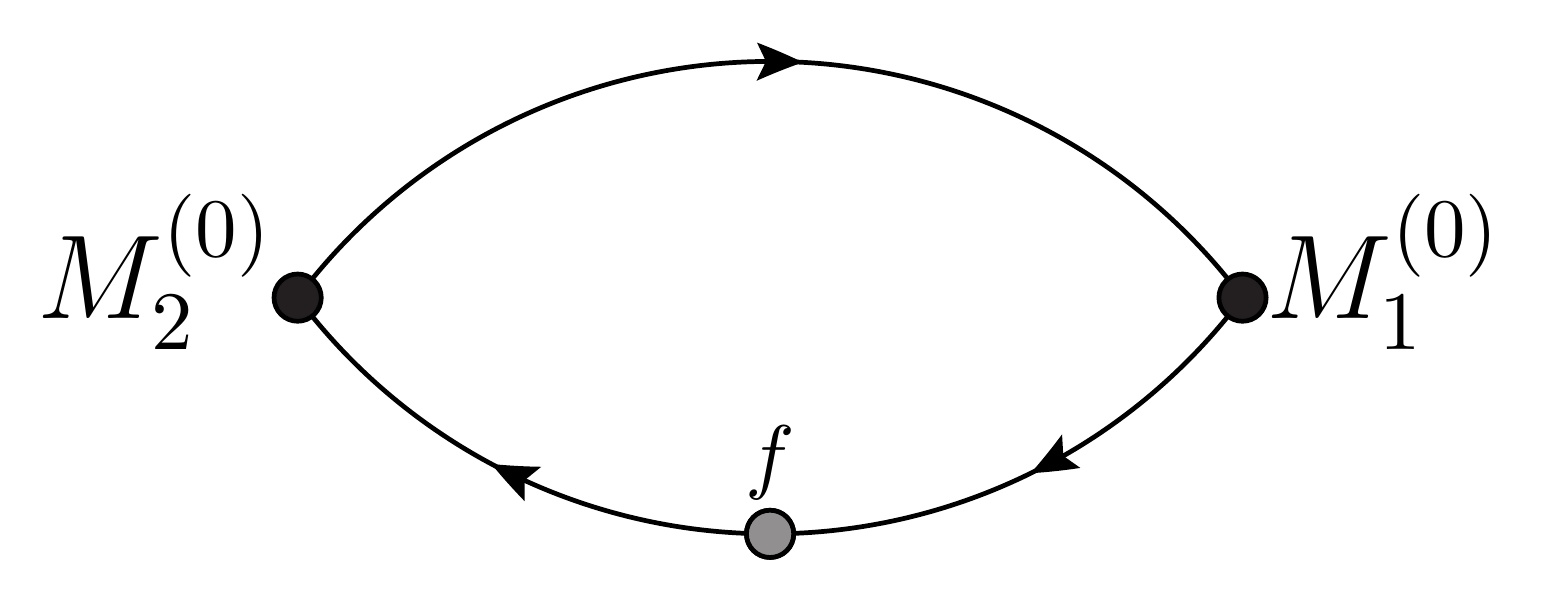}
\end{gathered}
+
\begin{gathered}
\includegraphics[width=6.5em]{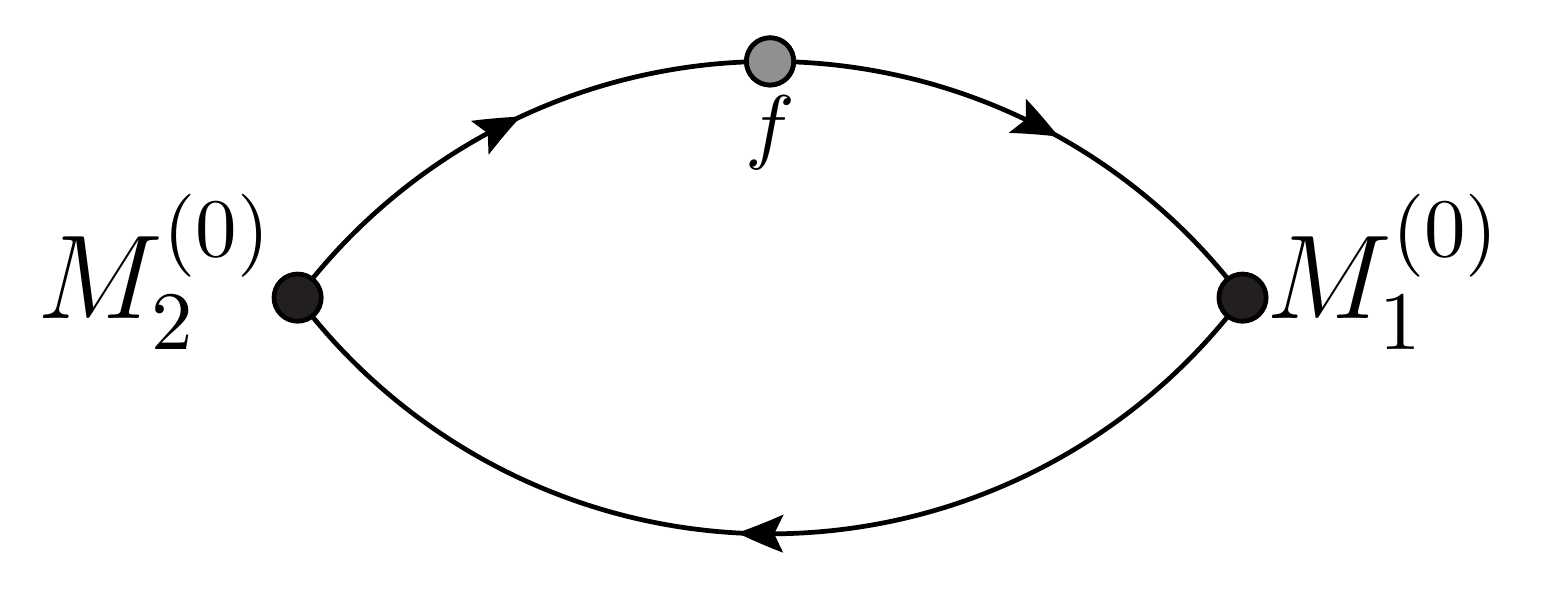}
\end{gathered}
\Big\rangle_{\mathrm{eff}}^{(0)}, \nonumber \\
C{}^{(1)}_{\Delta \beta} &= 
\begin{aligned}[t]
&\Big\langle
\begin{gathered}
\includegraphics[width=6.5em]{diagrams/mes2pt_con0.pdf}
\end{gathered}
\begin{gathered}
\includegraphics[width=1.5em]{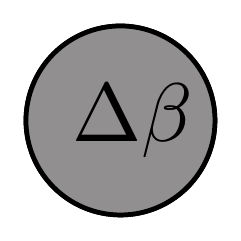}
\end{gathered}
\Big\rangle_{\mathrm{eff}}^{(0)} - \Big\langle
\begin{gathered}
\includegraphics[width=6.5em]{diagrams/mes2pt_con0.pdf}
\end{gathered}
\Big\rangle_{\mathrm{eff}}^{(0)}
\Big\langle
\begin{gathered}
\includegraphics[width=1.5em]{diagrams/vertex_beta.pdf}
\end{gathered}
\Big\rangle_{\mathrm{eff}}^{(0)},
\end{aligned} \nonumber \\
C{}^{(1)}_{e^{2}} &=
\Big\langle
\begin{aligned}[t]
&
\begin{gathered}
\includegraphics[width=6.5em]{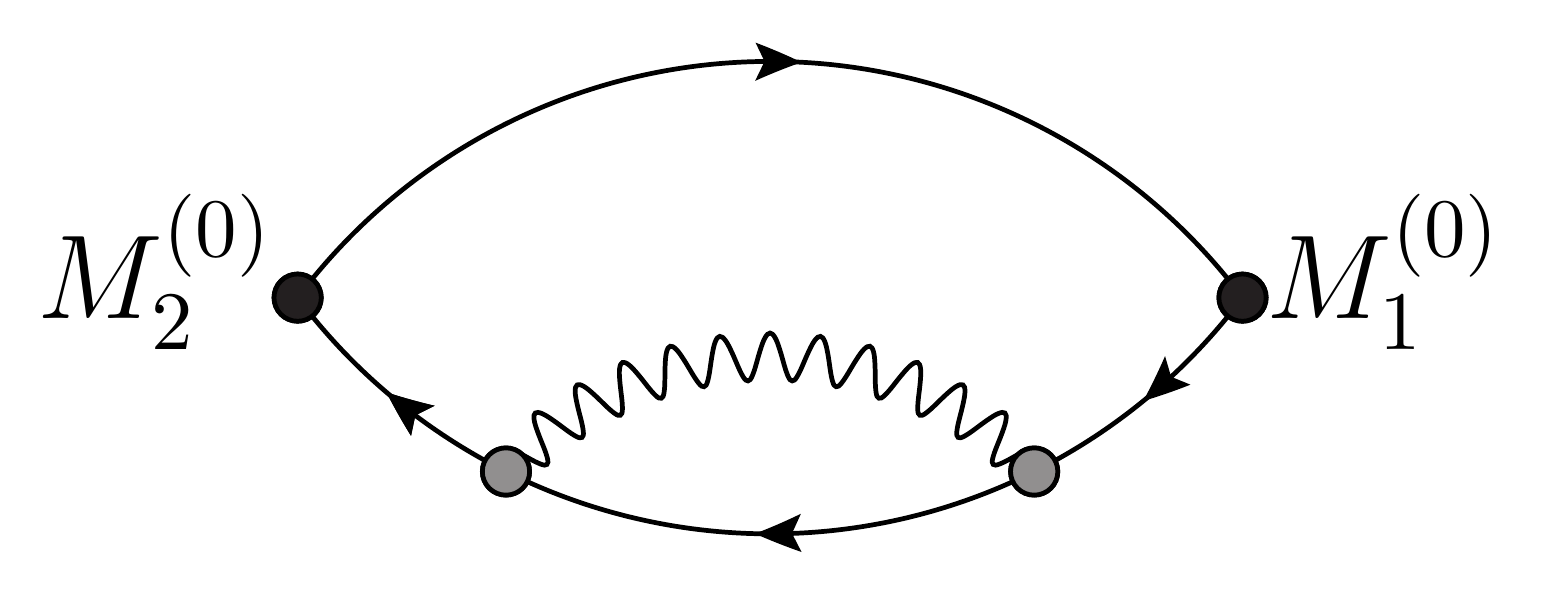}
\end{gathered}
+
\begin{gathered}
\includegraphics[width=6.5em]{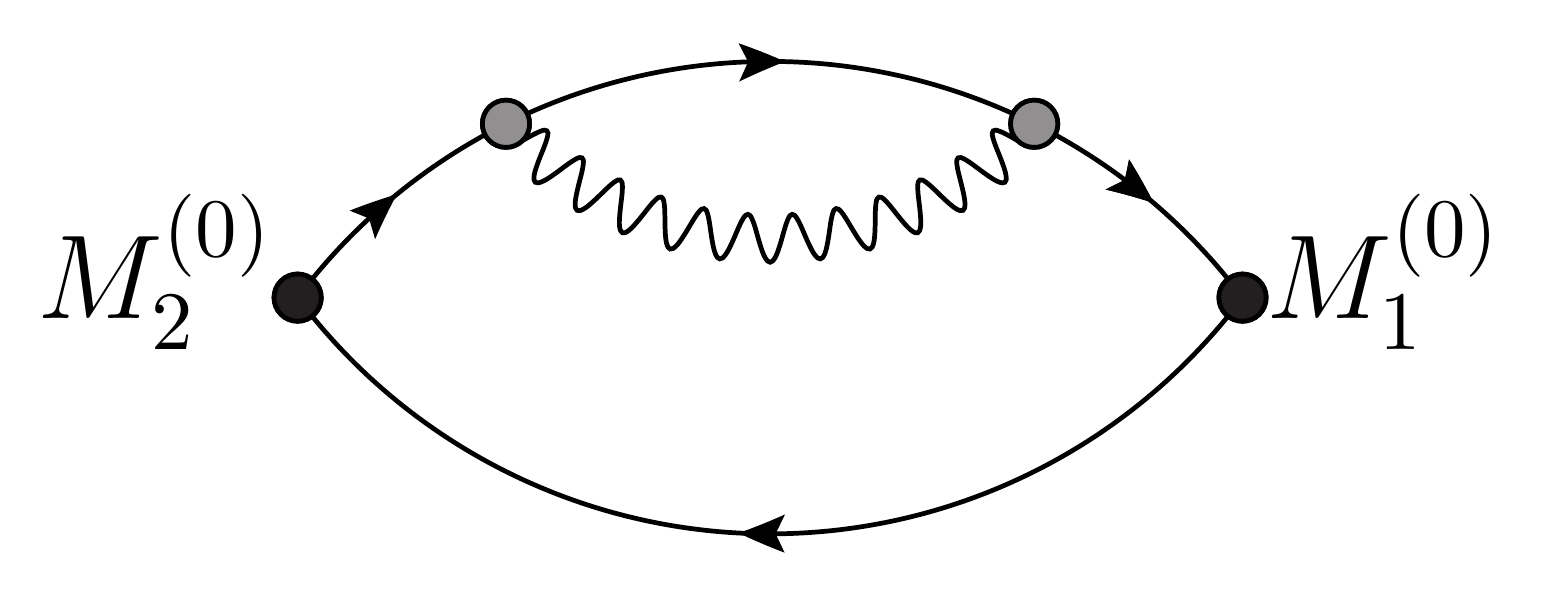}
\end{gathered}
+
\begin{gathered}
\includegraphics[width=6.5em]{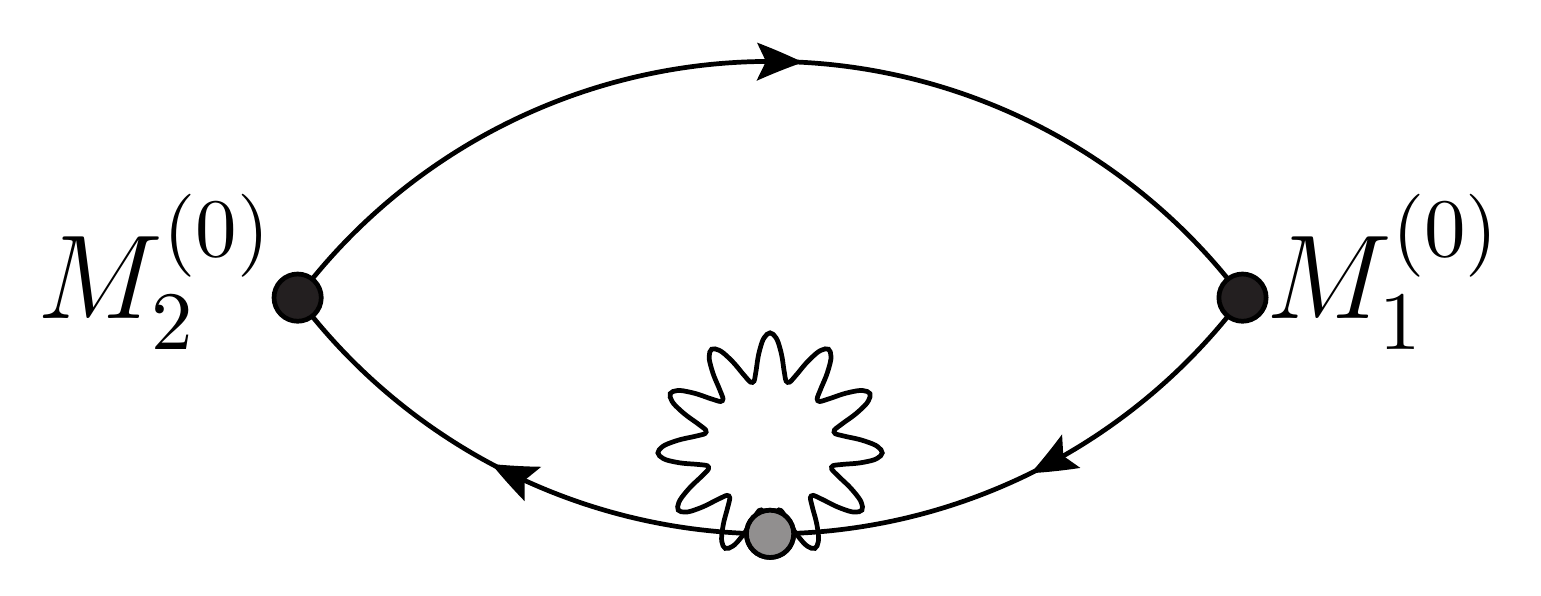}
\end{gathered}
+
\begin{gathered}
\includegraphics[width=6.5em]{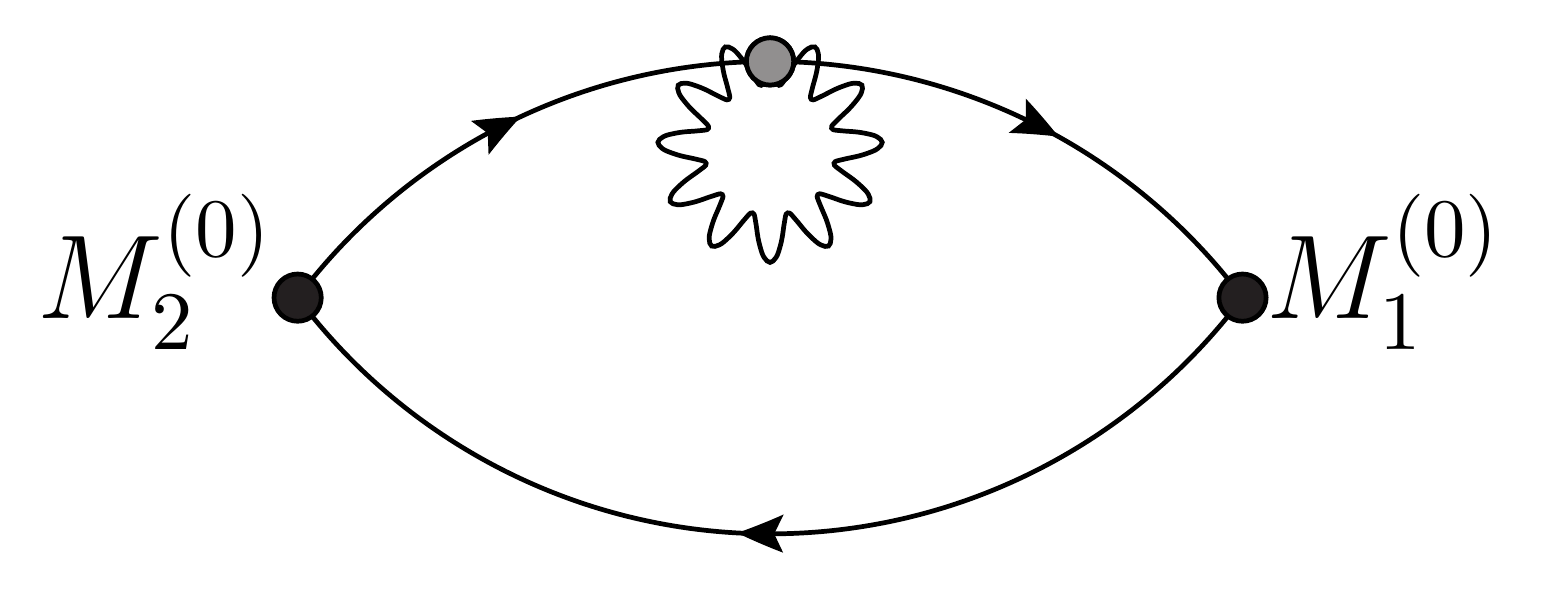}
\end{gathered} \\
&+
\begin{gathered}
\includegraphics[width=6.5em]{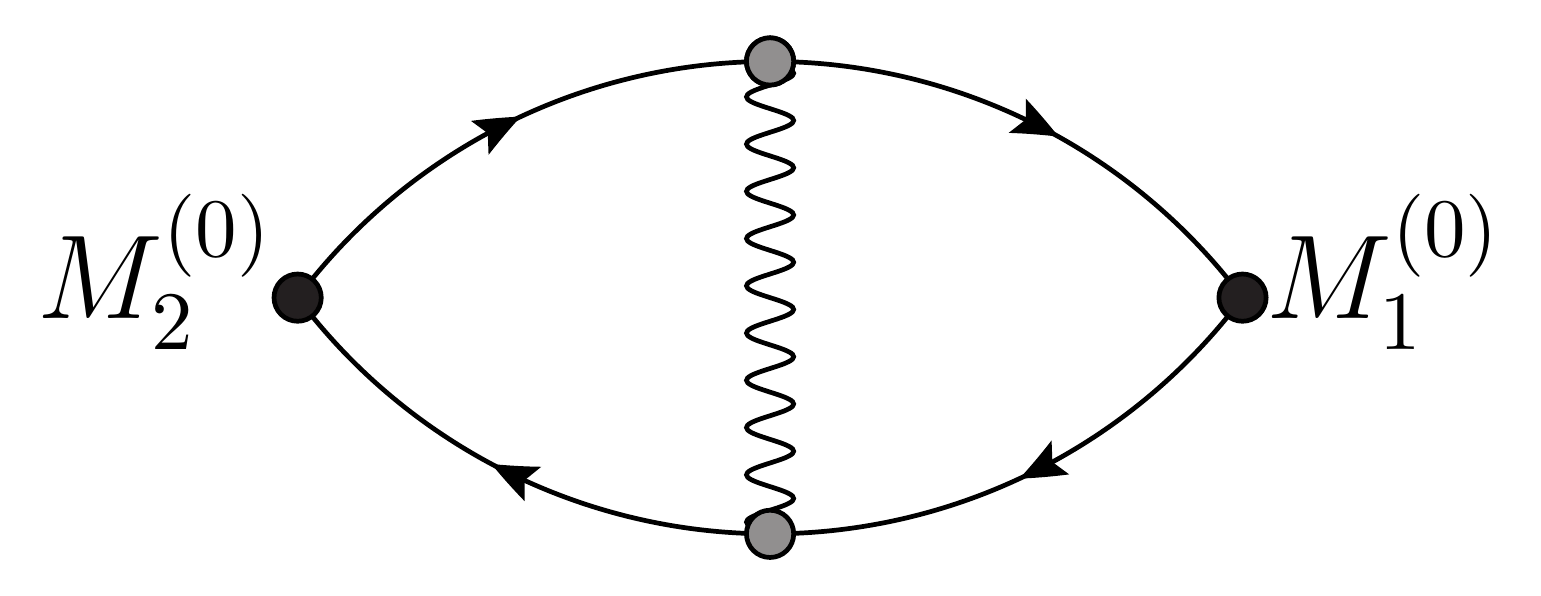}
\end{gathered}
+
\begin{gathered}
\includegraphics[width=6.5em]{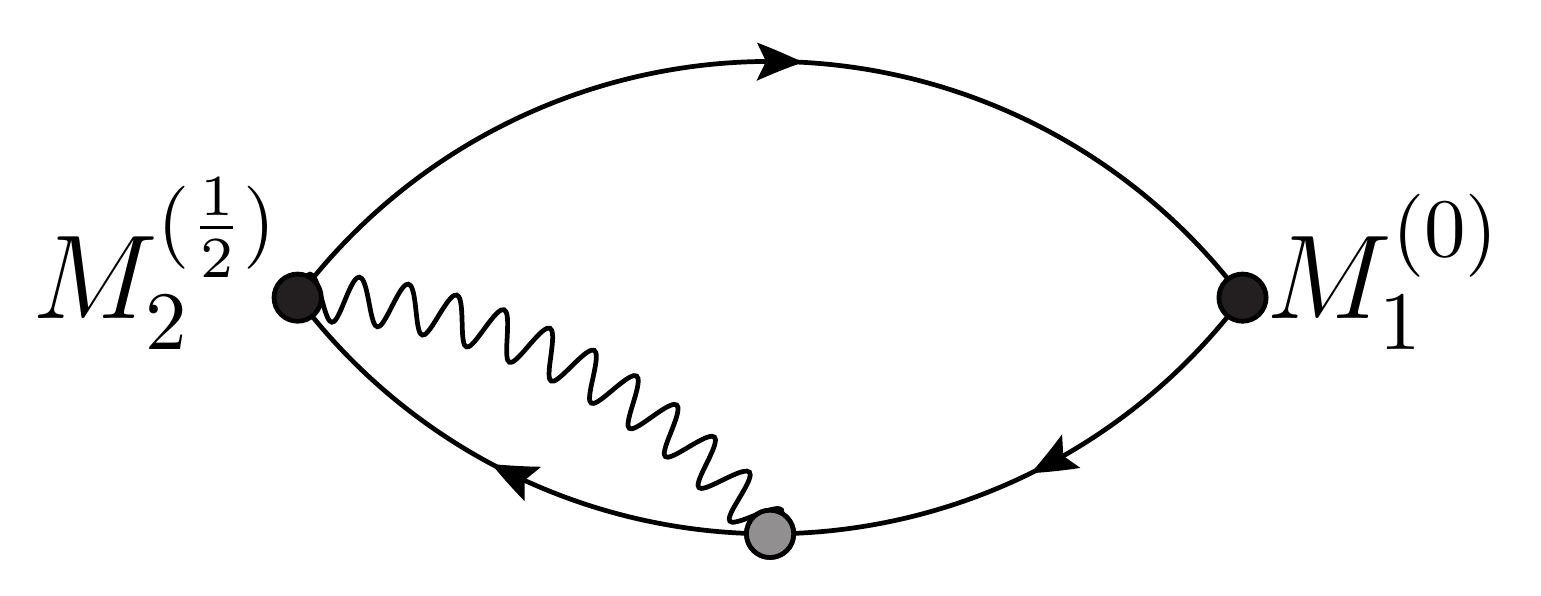}
\end{gathered}
+
\begin{gathered}
\includegraphics[width=6.5em]{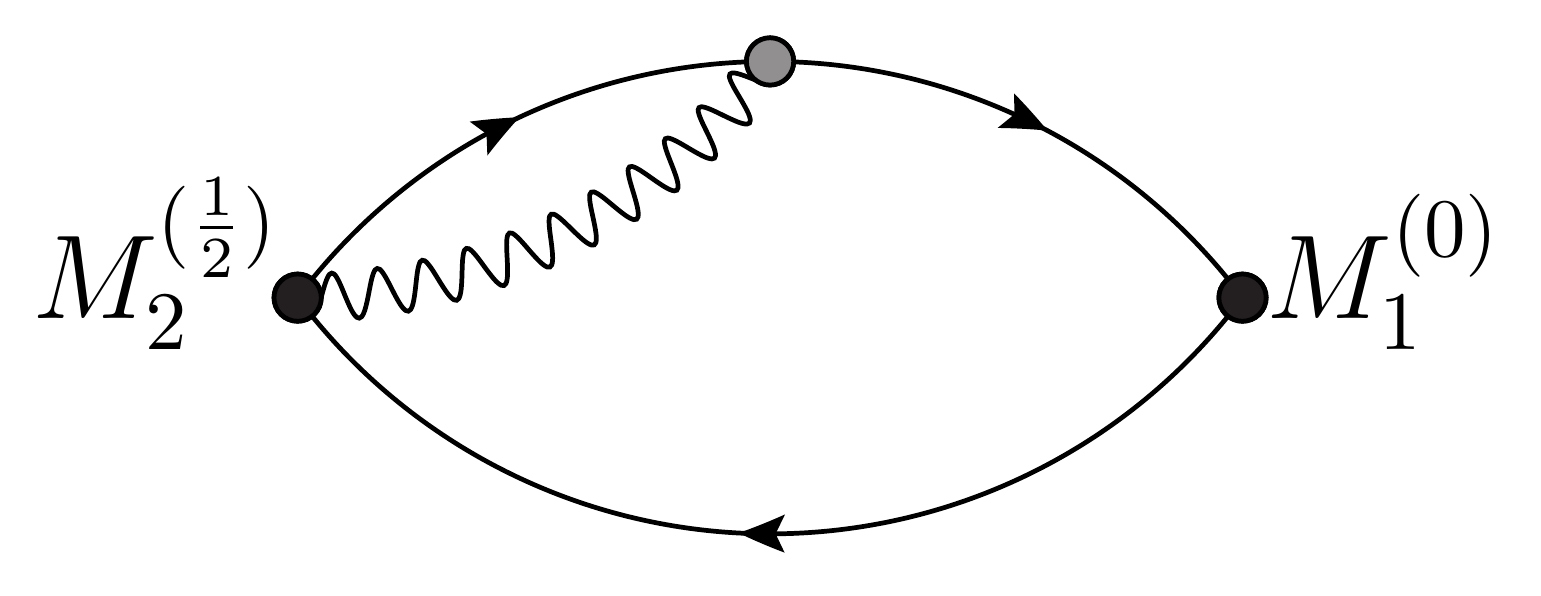}
\end{gathered}
+
\begin{gathered}
\includegraphics[width=6.5em]{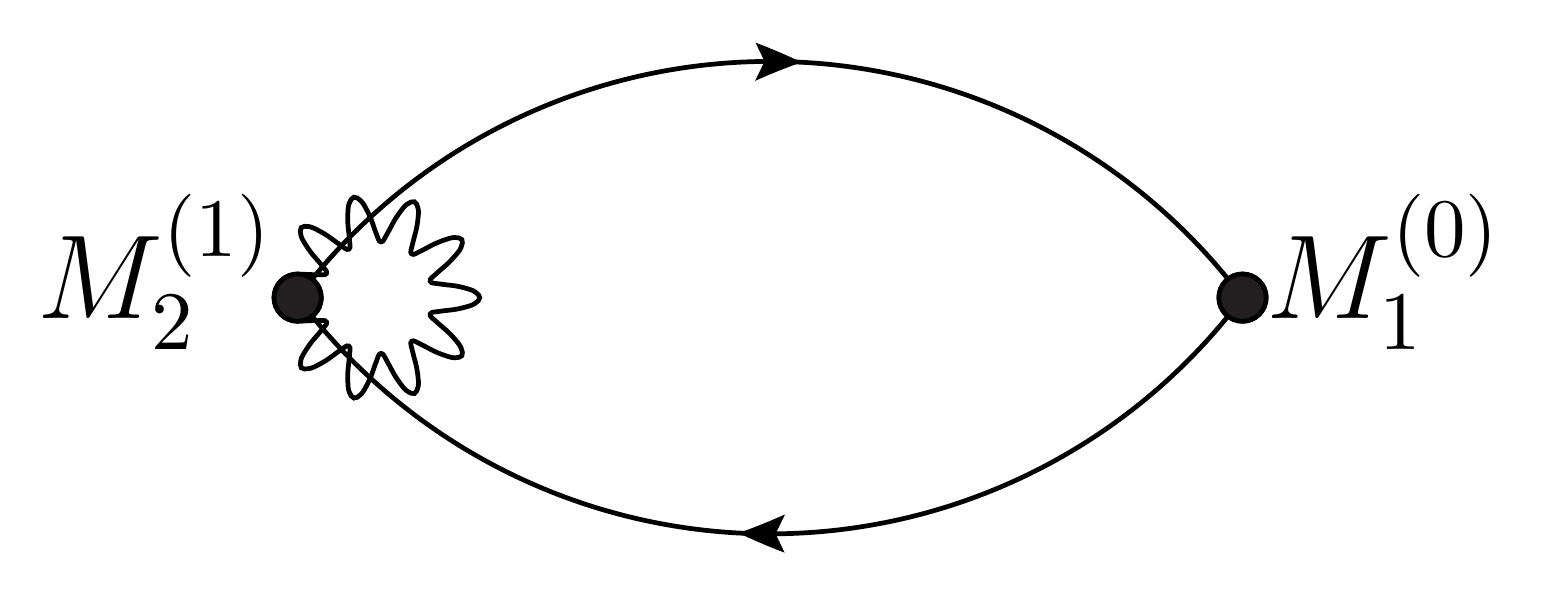}
\end{gathered}
\Big\rangle_{\mathrm{eff}}^{(0)}.
\end{aligned}
\label{eq:qconmes2pt}
\end{align}
We evaluate the diagrams by means of stochastic $U(1)$ quark sources with support on a single time-slice and $Z_{2}$ photon sources to estimate the all-to-all photon propagator in Coulomb gauge~\cite{Risch:2018ozp}. The photon boundary conditions are chosen in accordance with the gauge field boundary conditions of the QCD$_{\mathrm{iso}}$ ensembles. For temporal periodic gauge ensembles we use periodic boundary conditions for the photon field, whereas for open temporal boundary conditions we apply homogeneous Dirichlet and Neumann boundary conditions~\cite{Risch:2018ozp}. In order to reduce the stochastic noise we apply covariant approximation averaging~\cite{Shintani:2014vja} in combination with the truncated solver method~\cite{Bali:2009hu}. The simulation code is based on the QDP++~\cite{Edwards:2004sx} and FFTW3~\cite{FFTW05} libraries and the openQCD~\cite{Luscher:2014} framework. We have performed simulations on three gauge ensembles listed in \cref{table_lattice_parameters}.

\begin{table}
\begin{centering}
\begin{tabular}{|l|l|l|l|l|l|l|l|}
\hline
 & $(L/a)^3\times T/a$ & $a\,[\fm]$ & $m_{\pi}\,[\MeV]$ & $m_{K}\,[\MeV]$ & $m_{\pi}L$ & $L\,[\fm]$  & boundary \\
\hline  
N200 & $48^3\times128$ & $0.06426(76)$ & $282(3)$ & $463(5)$ & 4.4 & 3.1 & open \\  
D450 & $64^3\times128$ & $0.07634(97)$ & $217(3)$ & $476(6)$ & 5.4 & 4.9 & periodic \\
H102 & $32^3\times96$ & $0.08636(10)$ & $354(5)$ & $438(4)$ & 5.0 & 2.8 & open \\
\hline
\end{tabular}
\caption{Parameters of CLS ensembles with $N_{\mathrm{f}}=2+1$ quark flavours of non-perturbatively O(a) improved Wilson quarks and tree-level improved L\"uscher-Weisz gauge action~\cite{Bruno:2014jqa, Bruno:2016plf}.}
\label{table_lattice_parameters}
\end{centering}
\end{table}

\section{Renormalisation of the local vector current}

In QCD+QED the flavour-diagonal bare vector currents $\mathcal{V}_{d} = (\mathcal{V}^{0}_{d},\mathcal{V}^{3}_{d},\mathcal{V}^{8}_{d})$ with $\Lambda^{0} = \frac{1}{\sqrt{6}}\mathds{1}$, $\Lambda^{3} = \frac{1}{2}\lambda^{3}$ and $\Lambda^{8} = \frac{1}{2}\lambda^{8}$ for the local and conserved discretisations $d=\mathrm{l},\mathrm{c}$ may undergo mixing~\cite{Risch:2019xio}. We therefore introduce renormalisation factor matrices $Z_{\mathcal{V}_{d,\ren}\mathcal{V}_{d}}=\Big(Z_{\mathcal{V}_{d,\ren}^{i_{2}}\mathcal{V}_{d}^{i_{1}}}\Big)_{i_{2},i_{1}=0,3,8}$, such that the renormalised vector currents expressed in terms of the bare currents read $
\mathcal{V}_{d,\ren}= Z_{\mathcal{V}_{d,\ren}\mathcal{V}_{d}} \mathcal{V}_{d}$ for $d = \mathrm{l},\mathrm{c}$. For the conserved vector current $\mathcal{V}_{\mathrm{c}}$ we assume that mixing is absent and the renormalisation trivial due to the existence of a lattice vector Ward identity~\cite{Risch:2019xio,peskin1997introduction}, i.e. $Z_{\mathcal{V}_{\mathrm{c},\ren}\mathcal{V}_{\mathrm{c}}}=\mathds{1}$. For a critical account on this assumption we refer to~\cite{Collins:2005nj}. We impose the renormalisation condition~\cite{Maiani:1986yj} $\langle 0 | \mathcal{V}_{\mathrm{c},\ren} | V \rangle = \langle 0 | \mathcal{V}_{\mathrm{l},\ren} | V \rangle$ for a low-energy vector state $|V \rangle$. Defining the matrix of correlation functions $\langle \mathcal{V} \mathcal{V} \rangle = (\langle \mathcal{V}^{i_{2}} \mathcal{V}^{i_{1}} \rangle)_{i_{2},i_{1}=0,3,8}$ we may express this relation in terms of renormalised zero-momentum projected correlation functions:
\begin{align}
\langle \mathcal{V}_{\mathrm{c},\ren}^{x_{2}^{0}} \mathcal{V}_{\mathrm{l},\ren}^{x_{1}^{0}} \rangle &\rightarrow \langle \mathcal{V}_{\mathrm{l},\ren}^{x_{2}^{0}} \mathcal{V}_{\mathrm{l},\ren}^{x_{1}^{0}} \rangle \quad\text{for}\quad T\gg x_{2}^{0} \gg x_{1}^{0} \gg 0.
\end{align}
Using the renormalisation relation of the vector currents, $Z_{\mathcal{V}_{\mathrm{c},\ren}\mathcal{V}_{\mathrm{c}}}=\mathds{1}$ and multiplying by $(Z_{\mathcal{V}_{\mathrm{l},\ren}\mathcal{V}_{\mathrm{l}}})^{-1}$ from the right, this condition becomes~\cite{Risch:2019xio}
\begin{align}
\langle \mathcal{V}_{\mathrm{c}}^{x_{2}^{0}} \mathcal{V}_{\mathrm{l}}^{x_{1}^{0}} \rangle &\rightarrow Z_{\mathcal{V}_{\mathrm{l},\ren}\mathcal{V}_{\mathrm{l}}} \,\langle \mathcal{V}_{\mathrm{l}}^{x_{2}^{0}} \mathcal{V}_{\mathrm{l}}^{x_{1}^{0}} \rangle \quad\text{for}\quad T\gg x_{2}^{0} \gg x_{1}^{0} \gg 0.
\end{align}
Hence, we may extract the renormalisation factor matrix for the spatial vector currents from~\cite{Risch:2019xio}
\begin{align}
Z_{\mathrm{eff},\mathcal{V}_{\mathrm{l},\ren}\mathcal{V}_{\mathrm{l}}}(x_{2}^{0},x_{1}^{0}) &= \Bigg(\frac{1}{3}\sum_{\mu=1}^{3} \langle \mathcal{V}_{\mathrm{c}}^{x_{2}^{0}\mu} \mathcal{V}_{\mathrm{l}}^{x_{1}^{0}\mu} \rangle\Bigg)\,\Bigg(\frac{1}{3}\sum_{\mu=1}^{3}\langle \mathcal{V}_{\mathrm{l}}^{x_{2}^{0}\mu} \mathcal{V}_{\mathrm{l}}^{x_{1}^{0}\mu} \rangle\Bigg)^{-1},
\end{align}
which satisfies $Z_{\mathrm{eff},\mathcal{V}_{\mathrm{l},\ren}\mathcal{V}_{\mathrm{l}}}(x_{2}^{0},x_{1}^{0}) \rightarrow Z_{\mathcal{V}_{\mathrm{l},\ren}\mathcal{V}_{\mathrm{l}}}$ for $T\gg x_{2}^{0} \gg x_{1}^{0} \gg 0$ in the limit of large time separations, where lattice artefacts become small. We further perform a perturbative expansion $Z_{\mathcal{V}_{\ren}\mathcal{V}} = (Z_{\mathcal{V}_{\ren}\mathcal{V}})^{(0)} + \sum_{l}\Delta\varepsilon_{l} \,(Z_{\mathcal{V}_{\ren}\mathcal{V}})^{(1)}_{l} + O(\Delta\varepsilon^{2})$. For the results of the extracted renormalisation factors we refer to~\cite{Risch:2019xio}. From the bare currents $\mathcal{V}_{d}$ and the renormalisation factor matrix $Z_{\mathcal{V}_{d,\ren}\mathcal{V}_{d}}$ we construct the renormalised electromagnetic current $\mathcal{V}{}_{d,\ren}^{\gamma}$ defined as
\begin{align}
\mathcal{V}{}_{d,\ren}^{\gamma} &= \mathcal{V}{}_{d,\ren}^{3} + \frac{1}{\sqrt{3}} \mathcal{V}{}_{d,\ren}^{8} = \sum_{i=0,3,8} \Big(Z_{\mathcal{V}_{d,\ren}^{3}\mathcal{V}_{d}^{i}} + \frac{1}{\sqrt{3}} Z_{\mathcal{V}_{d,\ren}^{8}\mathcal{V}_{d}^{i}}\Big)\,\mathcal{V}{}_{d}^{i} \quad\text{for}\quad d = \mathrm{l},\mathrm{c}.
\end{align}

\section{The LO-HVP contribution to the muon anomalous magnetic moment $a_{\mu}$}
\label{sec:amu}
\begin{figure}
\centering
\includegraphics[width=0.49\textwidth]{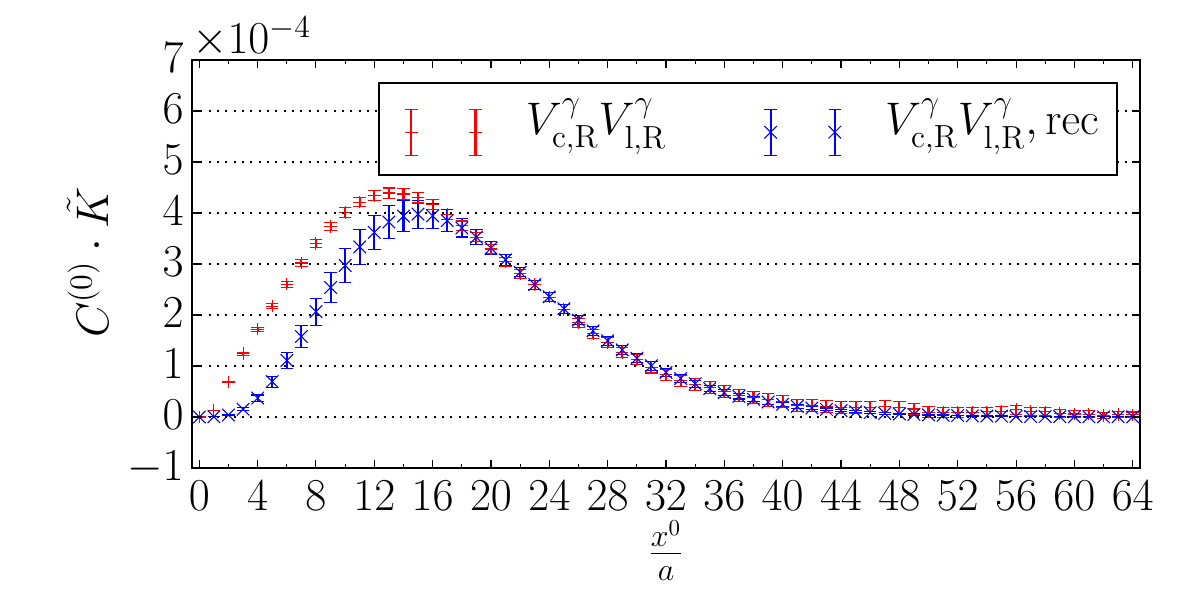}
\includegraphics[width=0.49\linewidth]{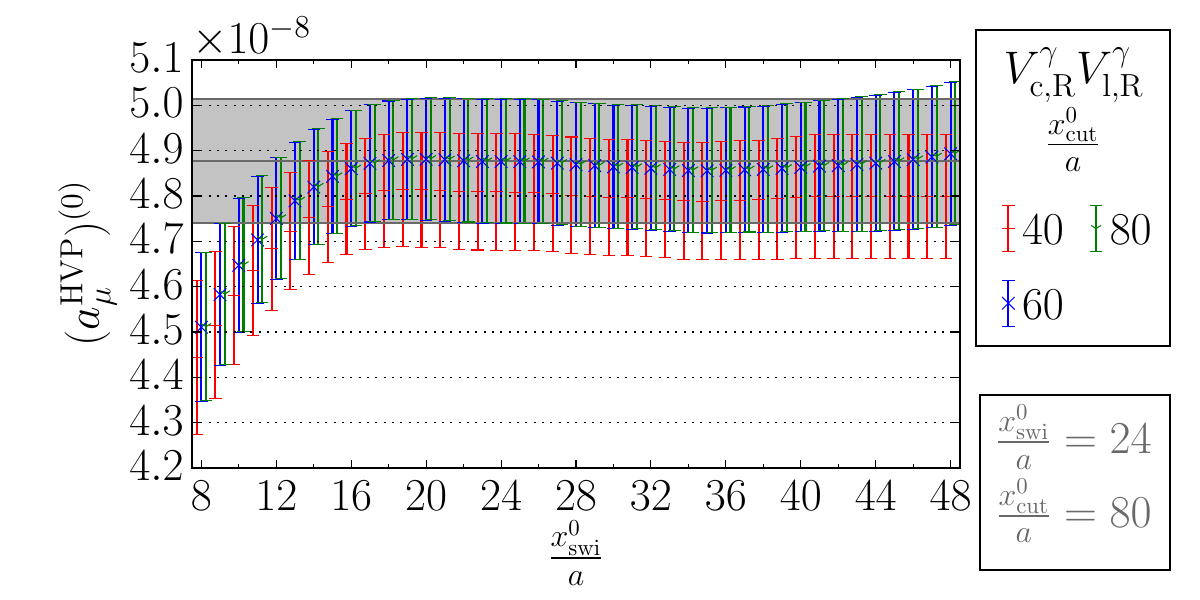}
\caption{Left: Integrand $\langle \mathcal{V}^{\gamma}_{\mathrm{c},\ren}\mathcal{V}^{\gamma}_{\mathrm{l},\ren} \rangle^{(0)}\cdot \widetilde{K}$ in red and its reconstruction in blue in lattice units on N200. $1\,\fm = 15.5(1)\,a$. Right: The corresponding $(a_{\mu}^{\mathrm{HVP}})^{(0)}$ as a function of $x^{0}_{\mathrm{swi}}$ and $x^{0}_{\mathrm{cut}}$.}
\label{fig:amuhvp0}
\end{figure}
\begin{figure}
\centering
\includegraphics[width=0.49\textwidth]{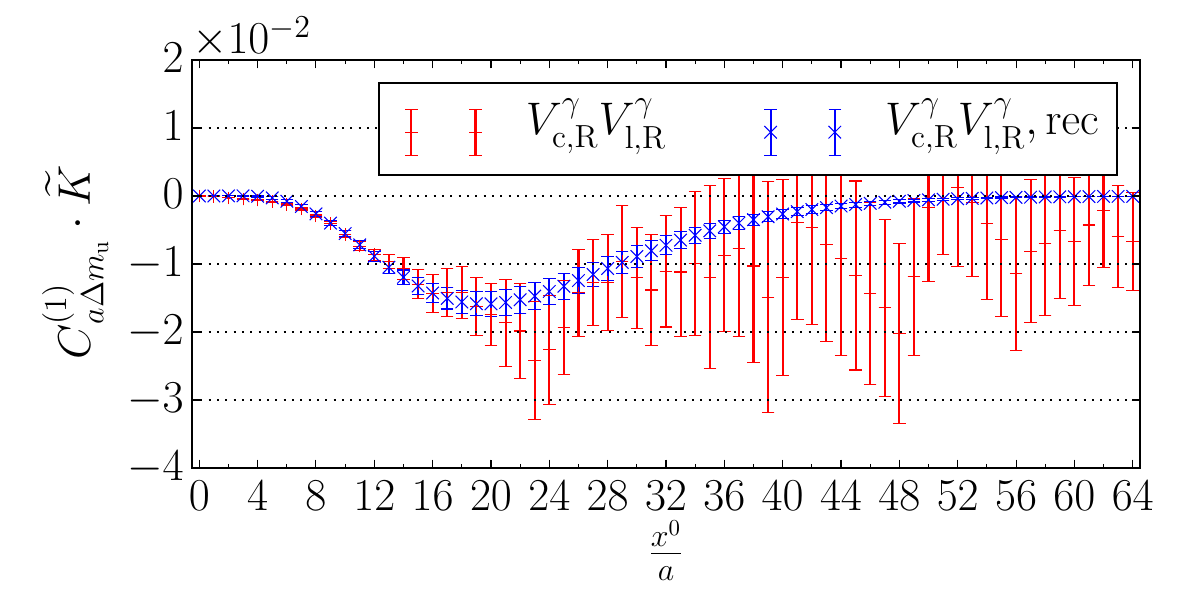}
\includegraphics[width=0.49\linewidth]{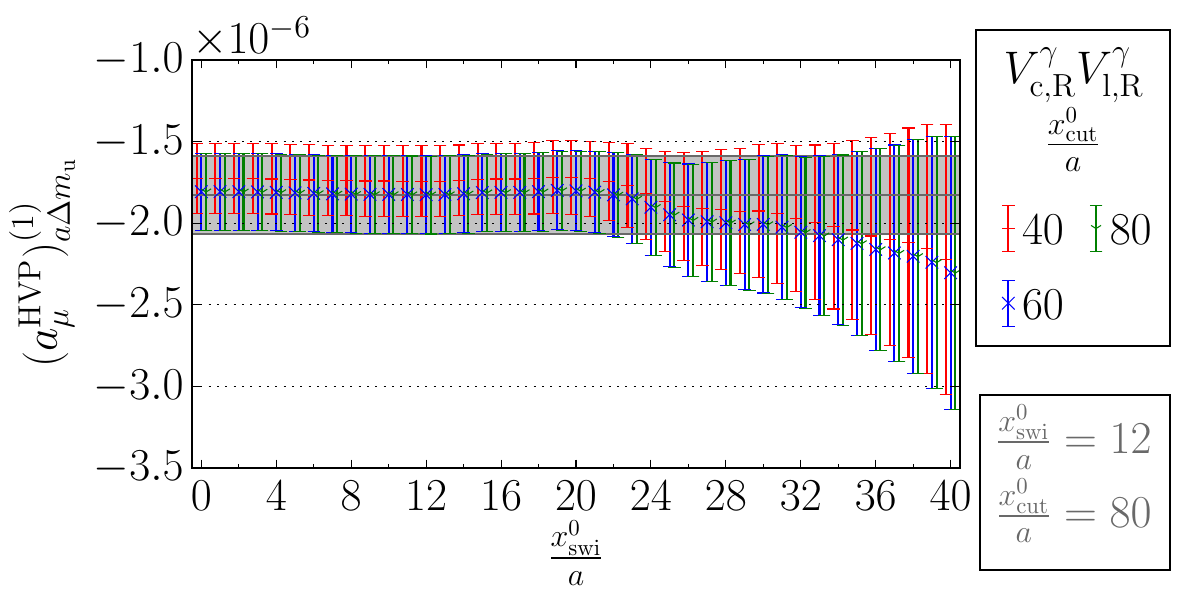}
\includegraphics[width=0.49\textwidth]{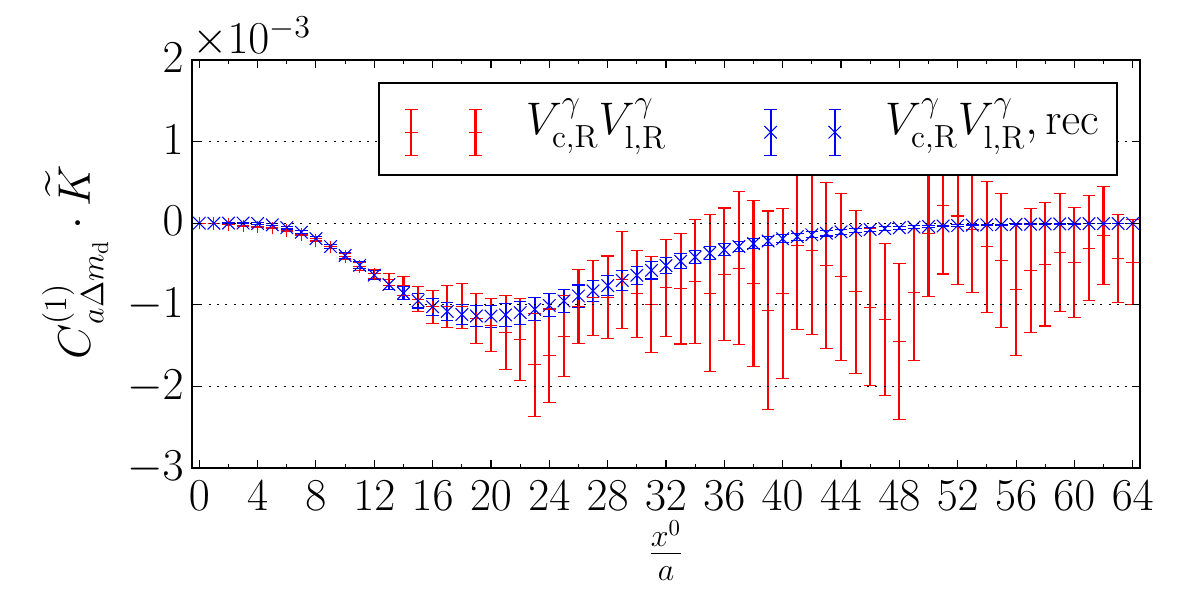}
\includegraphics[width=0.49\linewidth]{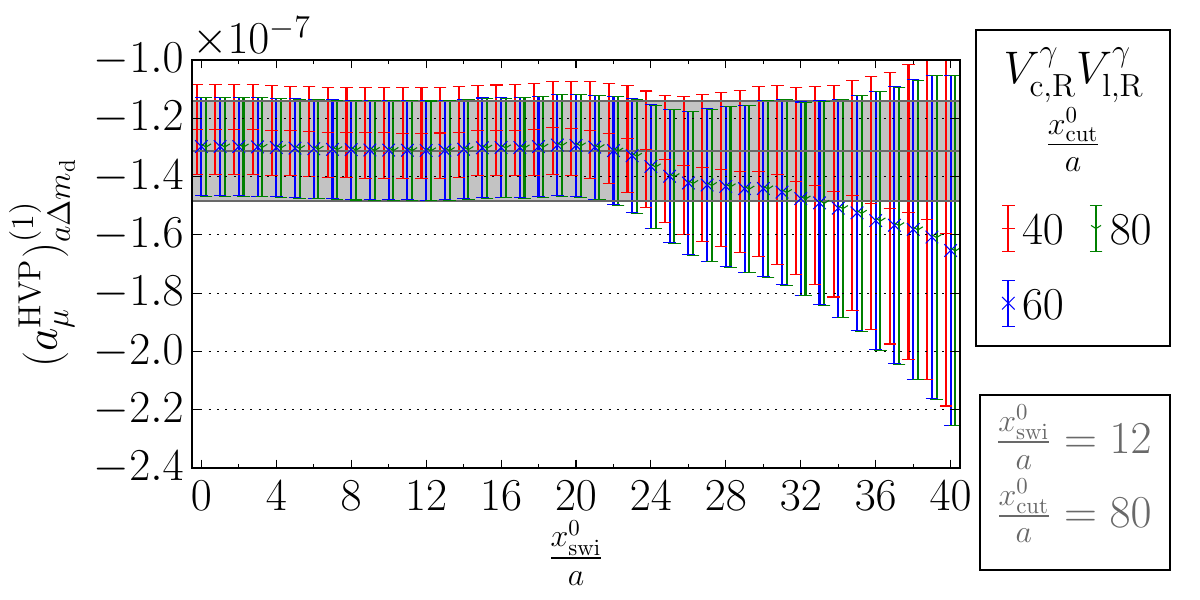}
\includegraphics[width=0.49\textwidth]{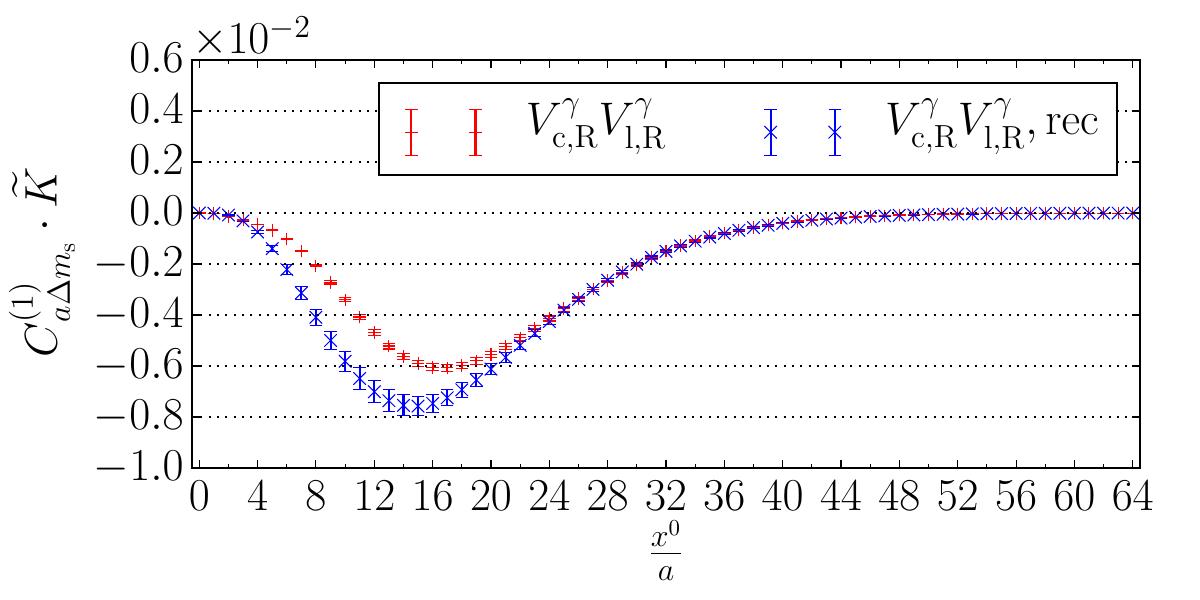}
\includegraphics[width=0.49\linewidth]{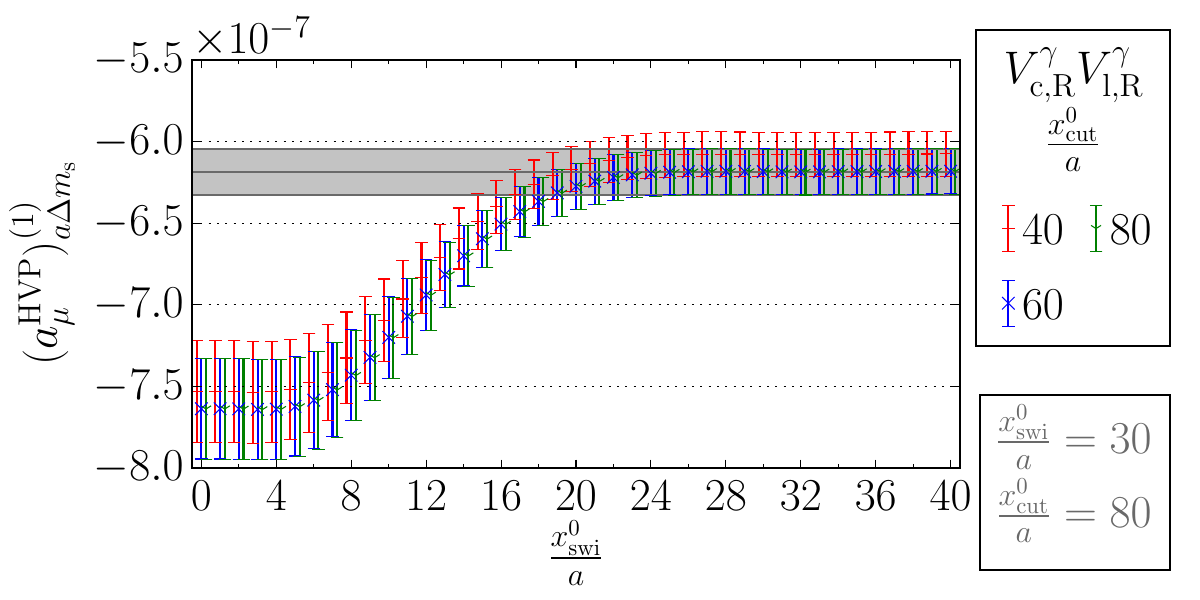}
\includegraphics[width=0.49\textwidth]{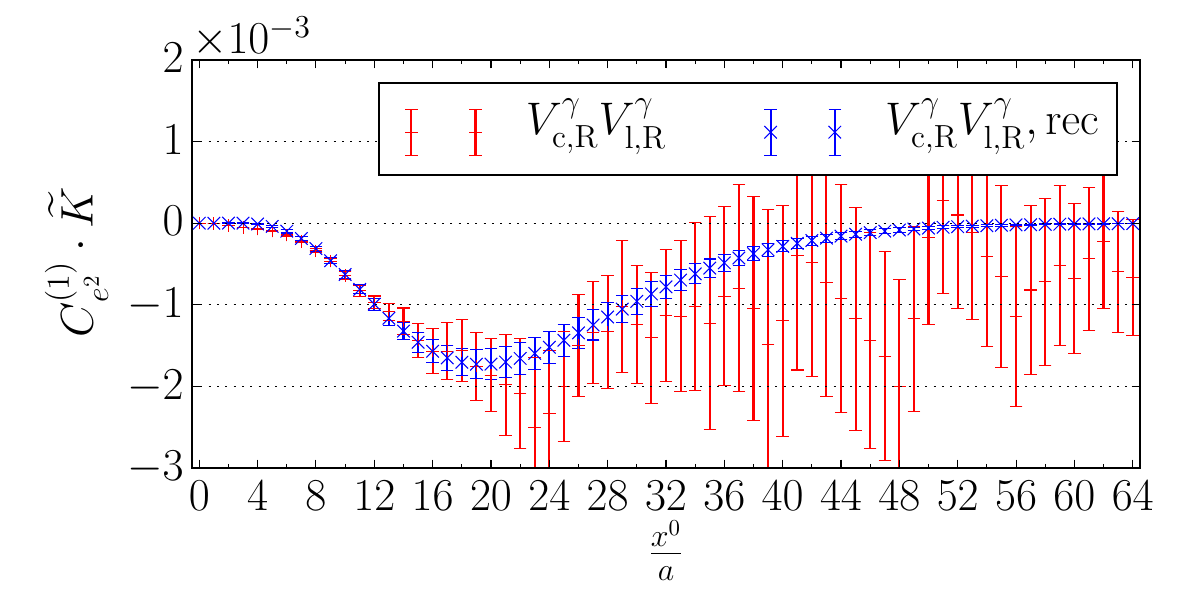}
\includegraphics[width=0.49\linewidth]{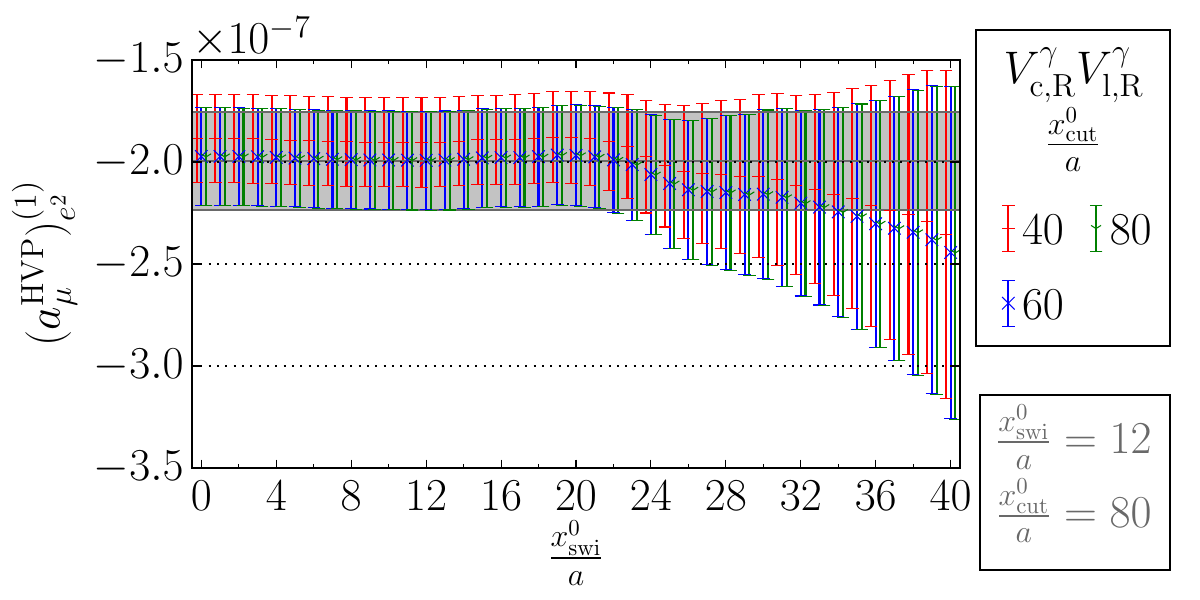}
\caption{Left: Integrand $\langle \mathcal{V}^{\gamma}_{\mathrm{c},\ren}\mathcal{V}^{\gamma}_{\mathrm{l},\ren}\rangle^{(1)}_{l}\cdot \widetilde{K}$ for $l=\dmu,\dmd,\dms,\ee$ in red and its reconstruction in blue in lattice units on N200. $1\,\fm = 15.5(1)\,a$. Right: The corresponding $(a_{\mu}^{\mathrm{HVP}})^{(1)}_{l}$ as a function of $x^{0}_{\mathrm{swi}}$ and $x^{0}_{\mathrm{cut}}$.}
\label{fig:amuhvp1}
\end{figure}
\begin{table}
\begin{subtable}[c]{\textwidth}
\begin{center}
\subcaption*{$a_{\mu}^{\mathrm{HVP}}$ from $\langle \mathcal{V}^{\gamma}_{\mathrm{c},\ren}\mathcal{V}^{\gamma}_{\mathrm{l},\ren} \rangle$}
\vspace{-0.2cm}
\begin{tabular}{|l|l|l|l|l|}
\hline
 & $(a_{\mu}^{\mathrm{HVP}})^{(0)}\,[10^{10}]$ & $(a_{\mu}^{\mathrm{HVP}})^{(1)}\,[10^{10}]$ & $a_{\mu}^{\mathrm{HVP}}\,[10^{10}]$ & $(a_{\mu}^{\mathrm{HVP}})^{(1)}/(a_{\mu}^{\mathrm{HVP}})^{(0)}$ \\
\hline
N200 & $488(9)_{\mathrm{{st}}}(10)_{\mathrm{{a}}}[14]$ & $-0.6[7]$ & $487(9)_{\mathrm{{st}}}(10)_{\mathrm{{a}}}[13]$ & $-0.0012[15]$ \\
D450 & $541(8)_{\mathrm{{st}}}(12)_{\mathrm{{a}}}[15]$ & $0.97[99]$ & $542(9)_{\mathrm{{st}}}(12)_{\mathrm{{a}}}[15]$ & $0.0018[18]$ \\
H102 & $440(4)_{\mathrm{{st}}}(10)_{\mathrm{{a}}}[10]$ & $1.7[4]$ & $441(4)_{\mathrm{{st}}}(10)_{\mathrm{{a}}}[11]$ & $0.0038[8]$ \\ 
\hline
\end{tabular}
\end{center}
\end{subtable}
\begin{subtable}[c]{\textwidth}
\begin{center}
\subcaption*{$a_{\mu}^{\mathrm{HVP}}$ from $\langle \mathcal{V}^{\gamma}_{\mathrm{l},\ren}\mathcal{V}^{\gamma}_{\mathrm{l},\ren} \rangle$}
\vspace{-0.2cm}
\begin{tabular}{|l|l|l|l|l|}
\hline
 & $(a_{\mu}^{\mathrm{HVP}})^{(0)}\,[10^{10}]$ & $(a_{\mu}^{\mathrm{HVP}})^{(1)}\,[10^{10}]$ & $a_{\mu}^{\mathrm{HVP}}\,[10^{10}]$ & $(a_{\mu}^{\mathrm{HVP}})^{(1)}/(a_{\mu}^{\mathrm{HVP}})^{(0)}$ \\
\hline
N200 & $491(8)_{\mathrm{{st}}}(11)_{\mathrm{{a}}}[13]$ & $-0.8[7]$ & $490(8)_{\mathrm{{st}}}(11)_{\mathrm{{a}}}[13]$ & $-0.0016[14]$ \\
D450 & $546(8)_{\mathrm{{st}}}(12)_{\mathrm{{a}}}[15]$ & $1.49[99]$ & $548(8)_{\mathrm{{st}}}(13)_{\mathrm{{a}}}[15]$ & $0.0027[18]$ \\
H102 & $445(4)_{\mathrm{{st}}}(10)_{\mathrm{{a}}}[10]$ & $1.6[4]$ & $447(4)_{\mathrm{{st}}}(10)_{\mathrm{{a}}}[11]$ & $0.0036[8]$ \\ 
\hline
\end{tabular}
\end{center}
\end{subtable}
\setlength{\belowcaptionskip}{-8pt}
\caption{Isosymmetric contribution $(a_{\mu}^{\mathrm{HVP}})^{(0)}$ and first-order correction $(a_{\mu}^{\mathrm{HVP}})^{(1)}$ of the hadronic vacuum polarisation contribution $a_{\mu}^{\mathrm{HVP}}$ from the two descretisations $\langle \mathcal{V}^{\gamma}_{\mathrm{c},\ren}\mathcal{V}^{\gamma}_{\mathrm{l},\ren} \rangle$ and $\langle \mathcal{V}^{\gamma}_{\mathrm{l},\ren}\mathcal{V}^{\gamma}_{\mathrm{l},\ren} \rangle$. The statistical and scale setting errors are labelled with "st" and "a", respectively.}
\label{tbl:amuhvp}
\end{table}
\setlength{\belowcaptionskip}{8pt}

In continuous Euclidean spacetime the LO-HVP contribution $a_{\mu}^{\mathrm{HVP}}$ can be computed from the QCD-connected part of the renormalised vector-vector correlation function by means of the time-momentum representation~\cite{Bernecker:2011gh,Francis:2013fzp,DellaMorte:2017dyu}
\begin{align}
a_{\mu}^{\mathrm{HVP}}\delta^{\mu_{2}\mu_{1}} &= \Big(\frac{\alpha}{\pi}\Big)^{2} \int_{0}^{\infty} \mathrm{d}x^{0}\, \widetilde{K}(x^{0},m_{\mu}) \int \mathrm{d}x^{3}\langle \mathcal{V}^{\gamma x\mu_{2}}_{\ren} \mathcal{V}^{\gamma 0\mu_{1}}_{\ren}\rangle_{\mathrm{QCD-con}}, \nonumber\\
\widetilde{K}(t,m_{\mu}) &= -8\pi^{2}\int_{0}^{\infty} \frac{\dif\omega}{\omega}\,\frac{1}{m_{\mu}^{2}}\hat{s}Z(\hat{s})^{3}\,\frac{1-\hat{s}Z(\hat{s})}{1+\hat{s}Z(\hat{s})^{2}}\,\Big(\omega^{2}t^{2}-4\sin^{2}\Big(\frac{\omega t}{2}\Big)\Big),
\end{align}
\vspace{-0.1em}where $\widetilde{K}(x^{0},m_{\mu})$ is the muon mass dependent integration kernel~\cite{DellaMorte:2017dyu}, $Z(\hat{s}) = -\frac{\hat{s}-\sqrt{\hat{s}^{2}+4\hat{s}}}{2\hat{s}}$ and $\hat{s} = \frac{\omega^{2}}{m_{\mu}^{2}}$. In the following, we drop the subscript "QCD-con" as we only consider quark-connected diagrams, c.f. \cref{eq:qconmes2pt}. Otherwise, the QCD-disconnected QED-connected part has to be subtracted by hand as it corresponds to a higher order HVP insertion~\cite{Chakraborty:2018iyb}. We discretise the continuum expression replacing the time integration by a finite summation up to $x^{0}_{\mathrm{cut}}$ and average over the three spatial components of the vector-vector correlation function:\vspace{-0.1em}
\begin{align}
a_{\mu}^{\mathrm{HVP}} &= \Big(\frac{\alpha}{\pi}\Big)^{2} a\sum_{x^{0}=0}^{x^{0}_{\mathrm{cut}}} \widetilde{K}(x^{0},m_{\mu}) \,\frac{1}{3}\sum_{\mu=1}^{3}\langle \mathcal{V}^{\gamma x^{0}\mu}_{\ren} \mathcal{V}^{\gamma 0\mu}_{\ren}\rangle.
\label{eq:amuhvp}
\end{align}
\vspace{-0.1em}We treat the noise problem of
$\langle \mathcal{V}_{\ren}^{\gamma x^{0}}\mathcal{V}_{\ren}^{\gamma 0} \rangle$ for large $x^{0}$ by performing a single-state reconstruction $\langle \mathcal{V}_{\ren}^{\gamma x^{0}}\mathcal{V}_{\ren}^{\gamma 0} \rangle_{\mathrm{rec}} = c \,e^{-m x^{0}}$, where the parameters $c$ and $m$ are determined by a fit. Nevertheless, this is only an effective description as we cannot resolve various low-energy states. We switch between $\langle \mathcal{V}_{\ren}^{\gamma x^{0}}\mathcal{V}_{\ren}^{\gamma 0} \rangle$ and the reconstruction $\langle \mathcal{V}_{\ren}^{\gamma x^{0}}\mathcal{V}_{\ren}^{\gamma 0} \rangle_{\mathrm{rec}}$ at $x^{0}_{\mathrm{swi}}$ above which the signal is lost. We perform a perturbative expansion, but neglect isospin breaking effects in the scale $a$ for $am_{\mu}^{\phys}$. \Cref{fig:amuhvp0,fig:amuhvp1} show the isosymmetric and first-order contributions to the integrand in \cref{eq:amuhvp} and the corresponding contributions to $a_{\mu}^{\mathrm{HVP}}$ for N200. The reconstruction is particularly relevant for suppressing the noise at large distances for the isosymmetric contribution, as well as for the first-order contributions with the expansion parameters $\dmu$, $\dmd$ and $\ee$. Results for the three investigated ensembles are displayed in \cref{tbl:amuhvp}. At the given level of statistical accuracy we find compatible results for both lattice discretisations. The scale setting uncertainty dominates the error of $(a_{\mu}^{\mathrm{HVP}})^{(0)}$ and $(a_{\mu}^{\mathrm{HVP}})^{(1)}$ is a correction to $(a_{\mu}^{\mathrm{HVP}})^{(0)}$ smaller than $O(0.5\%)$. The first-order correction $(a_{\mu}^{\mathrm{HVP}})^{(1)}$ is smaller than the error of $(a_{\mu}^{\mathrm{HVP}})^{(0)}$.

\section{The LO hadronic contribution to the running of $\alpha_{\mathrm{em}}$}

\setlength{\belowcaptionskip}{-6pt}
\begin{figure}
\centering
\includegraphics[width=0.49\textwidth]{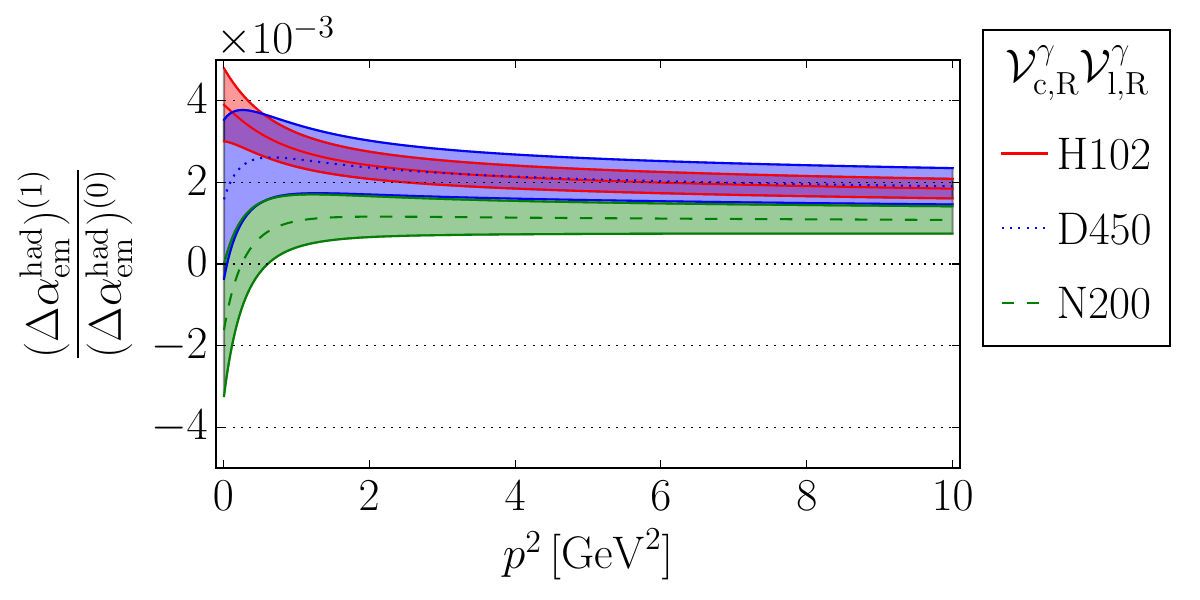}
\includegraphics[width=0.49\textwidth]{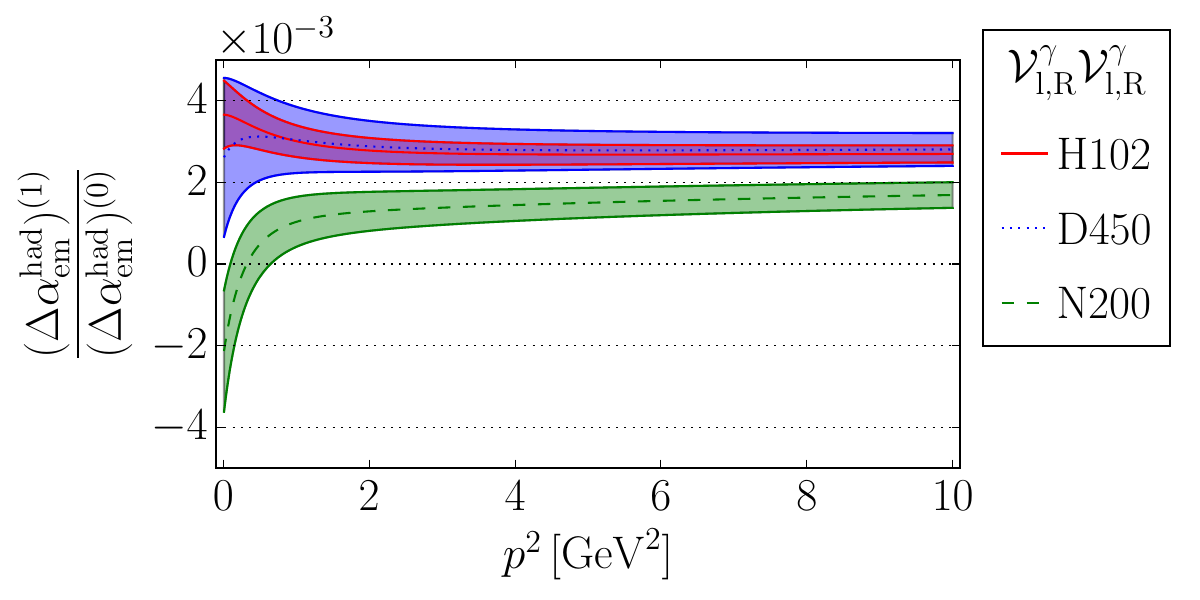}
\caption{Relative isospin breaking correction to the hadronic contributions $\Delta\alpha^{\mathrm{had}}_{\mathrm{em}}$ to the running of the electromagnetic coupling from the two descritisations $\langle \mathcal{V}^{\gamma}_{\mathrm{c},\ren}\mathcal{V}^{\gamma}_{\mathrm{l},\ren} \rangle$ and $\langle \mathcal{V}^{\gamma}_{\mathrm{l},\ren}\mathcal{V}^{\gamma}_{\mathrm{l},\ren} \rangle$.}
\label{fig:runningalpha}
\end{figure}
\setlength{\belowcaptionskip}{6pt}

The LO hadronic contribution to the running of $\alpha_{\mathrm{em}}$ is related to the subtracted hadronic vacuum polarisation function $\hat{\Pi}(p^{2}) = \Pi(p^{2}) - \Pi(0)$ by $\Delta\alpha_{\mathrm{em}}^{\mathrm{had}}(-p^{2}) = 4\pi\alpha_{\mathrm{em}}\,\hat{\Pi}_{\mathcal{V}^{\gamma}_{\ren}\mathcal{V}^{\gamma}_{\ren}}(p^{2})$. We compute $\hat{\Pi}$ in the time-momentum representation~\cite{Bernecker:2011gh}
\begin{align}
\hat{\Pi}_{\mathcal{V}^{\gamma}_{\ren}\mathcal{V}^{\gamma}_{\ren}}(p^2)\,\delta^{\mu_{2}\mu_{1}} &= \int_{0}^{\infty} \mathrm{d}x^{0}\, K(p^{2},x^{0})\int \mathrm{d}x^{3}\, \langle \mathcal{V}^{\gamma x\mu_{2}}_{\ren} \mathcal{V}^{\gamma 0\mu_{1}}_{\ren}\rangle_{\mathrm{QCD-con}}
\end{align}
with the kernel function $K(\omega^{2},t) = -\frac{1}{\omega^{2}}(\omega^{2}t^{2}-4\sin^{2}(\frac{\omega t}{2}))$. We treat $\langle \mathcal{V}^{\gamma}_{\ren}\mathcal{V}^{\gamma}_{\ren} \rangle$ as in \cref{sec:amu}. \Cref{fig:runningalpha} shows results for the quark-connected contributions depicted in \cref{eq:qconmes2pt}. For both discretisations we find corrections smaller than $O(0.5\%)$ on all investigated ensembles with the largest corrections in the small-momentum regime.

\section{Conclusions and Outlook}

We introduced a hadronic renormalisation scheme for QCD+QED and QCD$_{\text{iso}}$ inspired by chiral perturbation theory and computed leading isospin breaking effects in the LO-HVP contribution to the anomalous magnetic moment of the muon $a_{\mu}^{\mathrm{HVP}}$ as well as in the LO hadronic contributions to the running of the electromagnetic coupling $\Delta \alpha_{\mathrm{em}}^{\mathrm{had}}$. For both quantities we found corrections smaller than $O(0.5\%)$ on the investigated ensembles. In a similar fashion, leading isospin breaking effects can also be computed for hadronic contributions to the running of the weak mixing angle $\Delta\sin^{2}\Theta_{W}^{\mathrm{had}}$~\cite{SanJose:2021apl} based on the correlation function $\langle \mathcal{V}^{Z}_{\ren}\mathcal{V}^{\gamma}_{\ren} \rangle$, where $\mathcal{V}^{Z}_{\ren} = \mathcal{V}^{T_{3}}_{\ren} - \sin^{2}\Theta_{W}\mathcal{V}^{\gamma}_{\ren}$ and $\mathcal{V}^{T_{3}}_{\ren} = -\frac{1}{2\sqrt{6}}\mathcal{V}^{0}_{\ren}+\frac{1}{2}\mathcal{V}^{3}_{\ren}+\frac{1}{2\sqrt{3}}\mathcal{V}^{8}_{\ren}$. To incorporate isospin breaking effects in the scale setting, which has been neglected in this work, we have started to investigate masses of octet and decuplet baryons~\cite{Segner:2021yqo}. Due to the noise problem of the vector-vector correlation function we plan to base the renormalisation procedure of the local vector current on the vector Ward identity~\cite{Gerardin:2018kpy,Gerardin:2019rua}, which allows for a better controlled determination of the renormalisation factors. Additionally, we are aiming for the inclusion of leading order QED finite volume corrections for hadron masses~\cite{Borsanyi:2014jba} as well as for the HVP related observables~\cite{Bijnens:2019ejw}. 

\vspace{0.5em}
\begin{small}
We are grateful to our colleagues within the CLS initiative for sharing ensembles. Our calculations were performed on the HPC Cluster "Clover" at the Helmholtz Institute Mainz and on the HPC Cluster "Mogon II" at the University of Mainz. The authors gratefully acknowledge the Gauss Centre for Supercomputing e.V. (www.gauss-centre.eu) for funding this project by providing computing time on the GCS Supercomputer JUWELS at J{\"u}lich Supercomputing Centre (JSC) for project CHMZ21.
\end{small}
\vspace{-0.8em}
\bibliographystyle{JHEP}
\bibliography{proceedings.bib}

\end{document}